\documentclass[runningheads]{llncs}

\usepackage{graphicx}
\usepackage{amsmath}
\usepackage{amsfonts} 
\usepackage{hyperref}
\usepackage{cleveref}
\usepackage{booktabs}
\usepackage{multirow}
\usepackage{algorithm}
\usepackage{algorithmic}
\usepackage{graphicx}
\usepackage{subfig}
\usepackage{makecell}
\usepackage{pgfplots}
\usetikzlibrary{patterns}

\definecolor{DarkOrange}{RGB}{209,78,1}
\definecolor{Orange}{RGB}{255,127,14}
\definecolor{Green}{RGB}{65,197,114}
\definecolor{Blue}{RGB}{38,162,237}

\crefname{figure}{Fig.}{Figs.}
\crefname{table}{Table}{Tables}
\crefname{algorithm}{Algorithm}{Algorithms}
\crefname{equation}{Eq.}{Equations}
\crefname{section}{Section}{Sections}
\creflabelformat{equation}{#2#1#3}
\makeatletter
\def\endthebibliography{%
  \def\@noitemerr{\@latex@warning{Empty `thebibliography' environment}}%
  \endlist
}
\makeatother

\pgfplotsset{compat=newest,width=8cm,height=4.8cm,
        every axis/.append style={
            tick label style={font=\fontsize{6}{6.5}\selectfont},
            label style={font=\fontsize{6}{6.5}\selectfont}
        },
        legend image code/.code={
            \draw[mark repeat=2,mark phase=2]
                plot coordinates {
                    (0cm,0cm)
                    (0.15cm,0cm)        
                    (0.3cm,0cm)
                };
        },
        major tick length=0.03cm,
        xtick align=outside,ytick align=outside,
        axis x line*=bottom,axis y line*=left,axis line style=ultra thin
}

\begin{document}
\title{Real-time Air Pollution Prediction Based on Spatiotemporal Big Data}

\titlerunning{Real-time Air Pollution Prediction Based on Spatiotemporal Big Data}

\author{Van Duc Le \and Tien-Cuong Bui \and
Sang Kyun Cha}
\authorrunning{Le et al.}
\institute{Dept. of Electrical and Computer Engineering \\ Seoul National University, Seoul, South Korea \\ \email{\{levanduc,cuongbt91,chask\}@snu.ac.kr}}

\maketitle              % typeset the header of the contribution

\begin{abstract}
Air pollution is one of the most concerns for urban areas. Many countries have constructed monitoring stations to hourly collect pollution values. Recently, there is a research in Daegu city, Korea for real-time air quality monitoring via sensors installed on taxis running across the whole city. The collected data size is massive (1-second interval) and in both Spatial and Temporal format. In this paper, based on this spatiotemporal Big data, we propose a real-time air pollution prediction model based on the convolutional neural network algorithm for image-like Spatial distribution of air pollution. Regarding temporal information in the data, we introduce a combination of a long short-term Memory unit for time series data and a neural network model for other air pollution impact factors such as weather conditions to build a hybrid prediction model. This model is simple in architecture, yet achieving good prediction ability.

\end{abstract}

\section{Introduction}
Outdoor air pollution is now seriously threatening to the human health and life in big cities, especially for elderly and children [1, 2]. The pollution data are usually collected hourly, which makes the air pollution prediction occur only by hour. A recent research in Daegu city, Korea tried to collect air pollution data in real-time. They installed air quality sensing devices on taxis running across Daegu city and collected sensor data every 1 second [9]. The dataset includes information of air pollutants and weather conditions, such as temperature, humidity, and air pressure. This is a Big data of air pollution dataset and we must use some big data computing engines such as Spark to pre-process data before using in later useful analysis. One of our proposals is a spatial analysis of air pollution levels. To show this spatial distribution clearly, the whole city map of Daegu is represented by a grid-map of e.g. 32x32. The air pollution values at each grid-cell are aggregated, and we represent this grid-map as a ``grayscale image'' of the same dimension 32x32. In Fig. 1, we can see the ``image'' of PM2.5 air pollution that spreads location by location and time by time across Daegu city. We have already known that the Convolutional Neural Network (CNN) algorithm was applied very successfully for Image classification [4]. In this paper, we also use a CNN model for ``image-like'' spatial distribution of air pollution to predict air pollution values in real-time.

\begin{figure}[ht]
    \centering
    % First Subfigure
    \subfloat[]{%
        \includegraphics[width=0.45\textwidth]{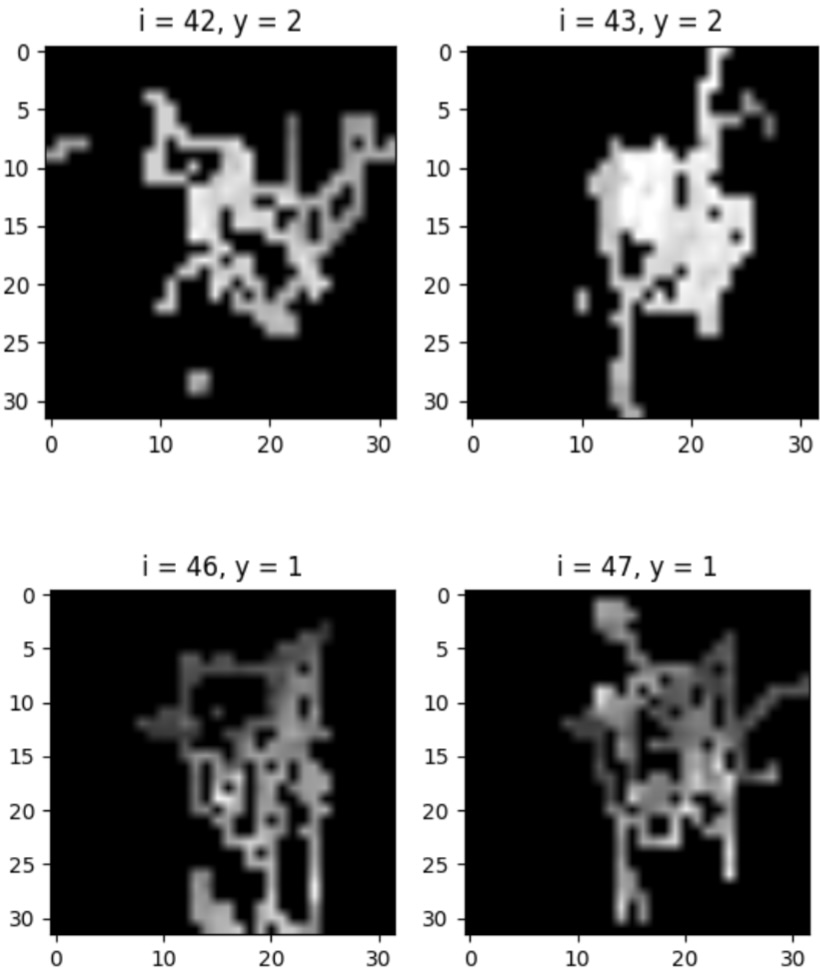}%
        \label{fig:sub1}%
    }
    \hfill 
    % Second Subfigure
    \subfloat[]{%
        \includegraphics[width=0.45\textwidth]{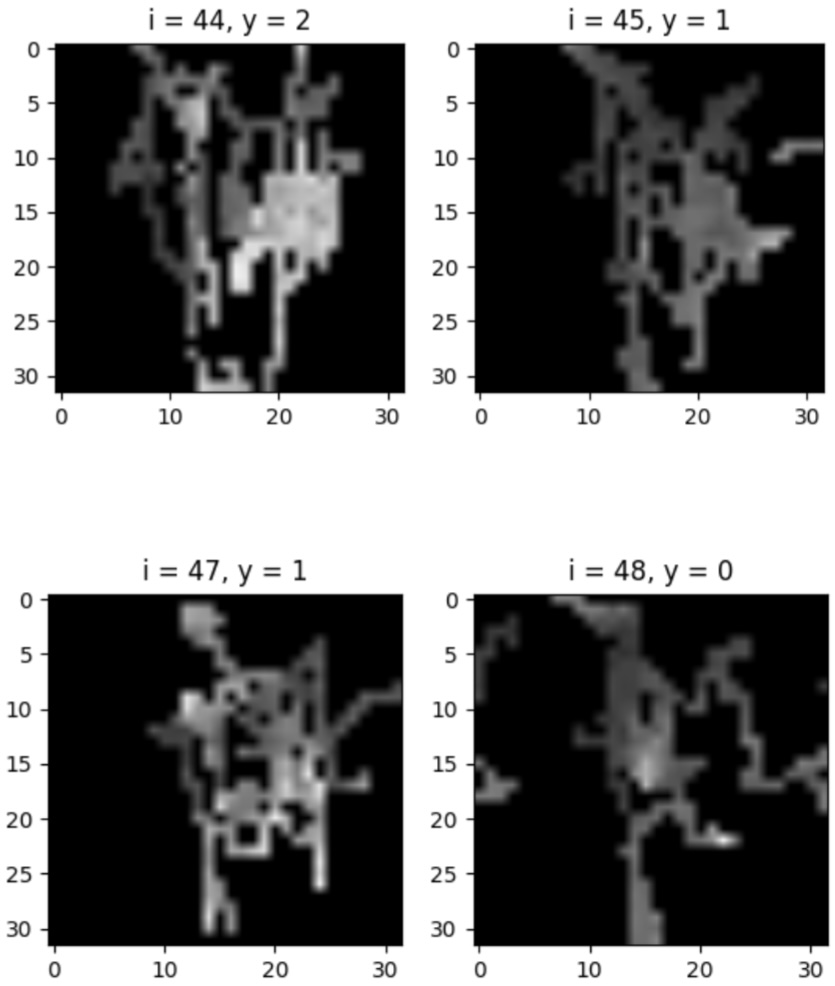}%
        \label{fig:sub2}%
    }
    
    \caption{``Image-like'' spatial distribution of PM2.5 air pollution values in Daegu city, Korea. $y = 2$ means the air pollution is Unhealthy, $y = 1$ means the pollution is Moderate, $y = 0$ means outdoor air is Good for health. The index $i$ indicates different time stamps.
}
    \label{fig:main}
\end{figure}

Air pollution is impacted by local meteorological factors, such as temperature, humidity levels, amounts of rain, wind (speed and direction) [3]. To affirm this statement, we gathered weather data, which was also collected in real-time by sensors in Daegu and analyzed them along with other information about air pollutants. In section 2.3, we show how we used a combination of Long-short Term Memory (LSTM) and a Neural Network (NN) to evaluate the relationship between these factors.
There are a number of papers which also leveraged the using of deep learning for air pollution prediction [5, 6]. Nevertheless, they all used hourly collected data by monitoring stations that are sparse and cannot represent well for the whole city. Moreover, most proposed models are the complicated combination of pollution data and other impacted factors such as weather, which makes it difficult to reveal changes in pollution levels and effects of other factors.
Our contributions have 2 main points:
\begin{enumerate}
    \item We apply a CNN algorithm to the ``image-like'' air pollution distribution to predict air pollution values in real-time.
    \item We propose a hybrid model of the LSTM algorithm for time series data and a neural network for other influential factors of air pollution levels. The architecture is simple, yet produces accurate air pollution prediction and unveils the effect of other factors to air pollution's change.
\end{enumerate}

\section{Proposed System}
\subsection{The real-time air pollution big data sensor dataset}

First, we elaborate on the procedure for collecting real-time sensor data in Daegu city. This is a project from the Korea Institute of Science and Technology Information (KISTI) and they wanted to create a Mobile Urban Sensing Dataset. They designed a circuit board etching many air pollutants sensors along with weather sensors and then installed them on taxis running inside Daegu city (Fig. 2). The 1-second interval data from sensors of all running taxis are collected and stored in a MySQL database with approximate 33.3 million rows so far (from June 2017 to March 2018) [8, 10]. Because of the large data volume, we used Spark \textendash a big data computing engine to pre-process data and transform them into inputs for further data analysis steps, as described in the next sections.

\begin{figure}[ht]
    \centering
    % First Subfigure
    \subfloat[Air quality and weather sensors]{%
        \includegraphics[width=0.45\textwidth]{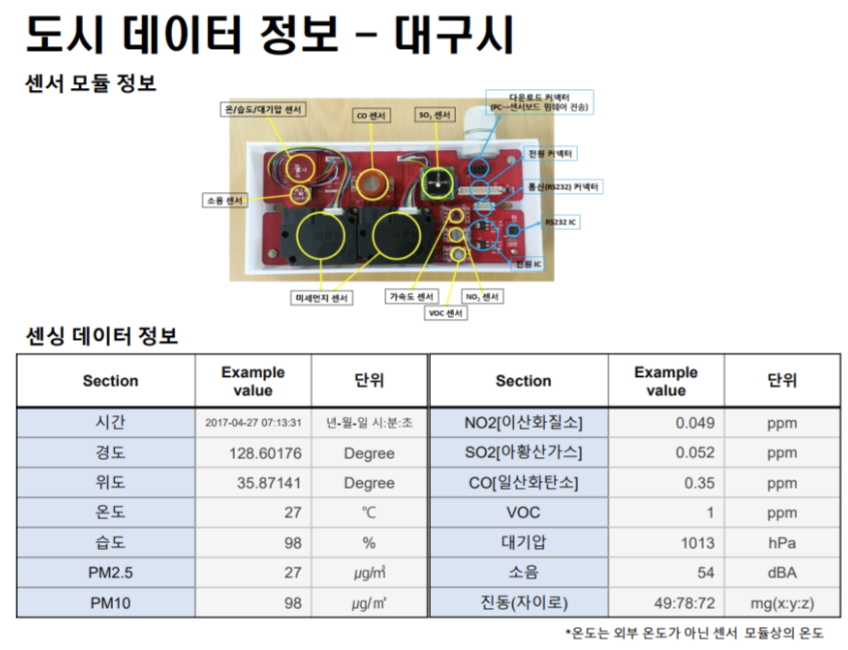}%
        \label{fig:sub1}%
    }
    \hfill 
    % Second Subfigure
    \subfloat[Real-time data Web UI]{%
        \includegraphics[width=0.45\textwidth]{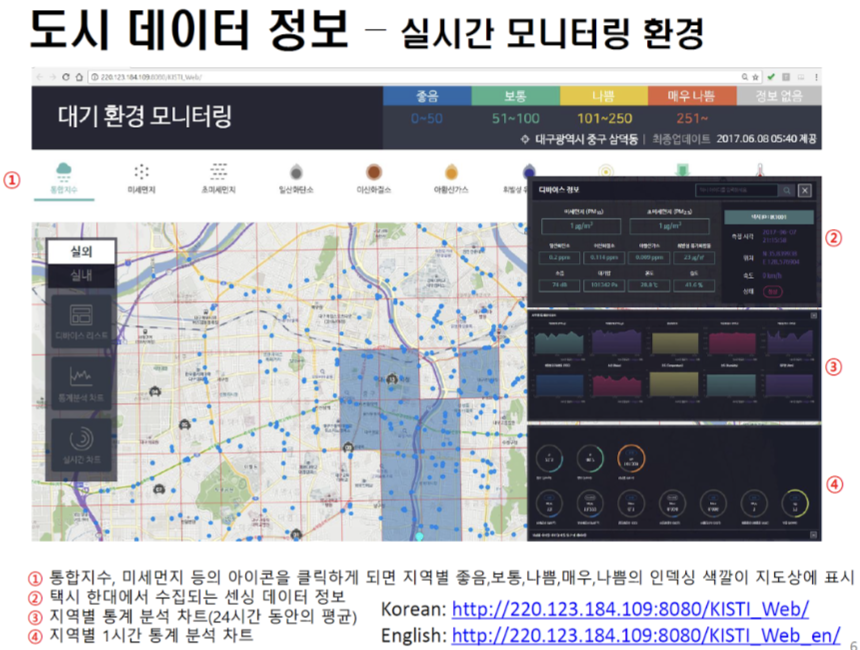}%
        \label{fig:sub2}%
    }
    
    \caption{Monitoring device information and Web UI of air quality data collection in Daegu city. Pictures are taken from [8].}
    \label{fig:main}
\end{figure}

\subsection{CNN for predicting real-time air pollution values}

\begin{figure}[ht]
    \centering
    \includegraphics[width=0.8\linewidth]{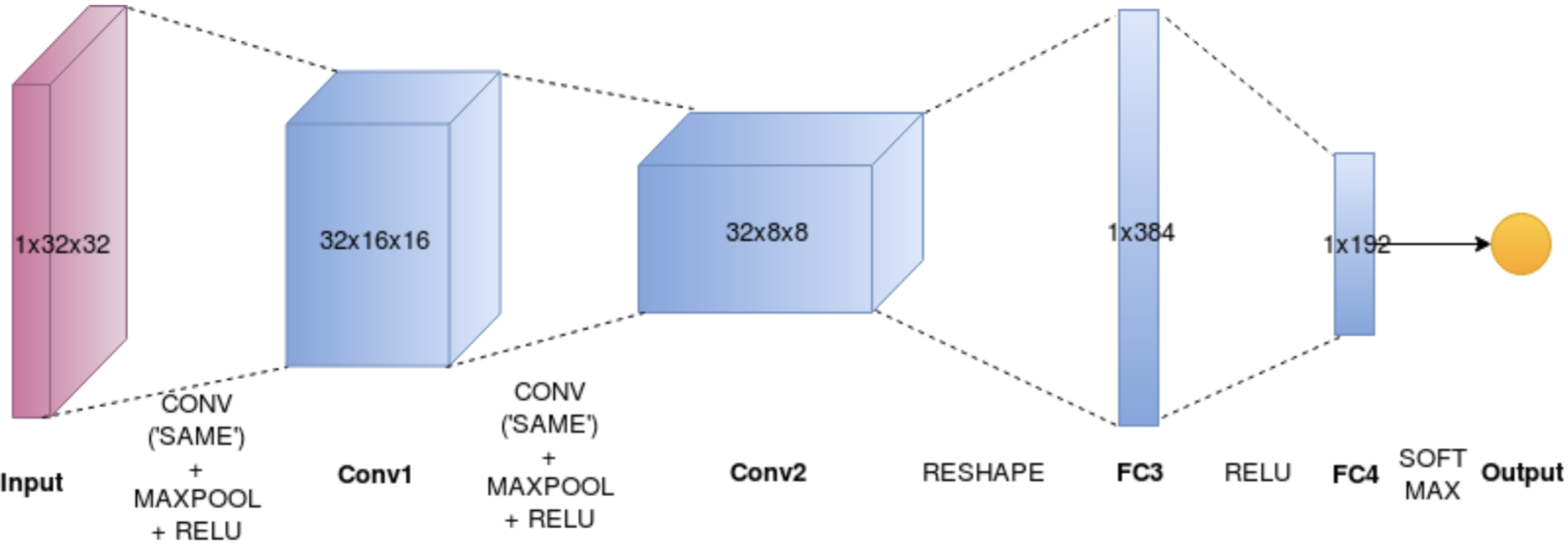}
    \caption{CNN architecture for real-time air pollution classification in Daegu.}
    \label{fig:cnn_arch}
\end{figure}

As in Fig. 1, we represent the distributed air pollutant values as a gray-scale image. The point is brighter, the air at that point (region) is more polluted and vice versa. This observation inspires us to use a CNN algorithm to predict air pollution value by classifying it as one of some states such as Good (for health), Moderate, or Unhealthy. Our model architecture derives from a CNN architecture that comprises of two consecutive (convolution + max pooling) layers followed by 2 fully-connected layers and a softmax layer.

\subsection{The combination of LSTM and NN to forecast air quality levels}
The air pollution values are time series data. Recently, the LSTM algorithm is popularly used for time series data prediction due to its capability to remember long-term sequences [5]. Similarly, we also use an LSTM in making a prediction model for air pollution values in some time period ahead (e.g., 8/12/24 hours). Air pollution values are aggregated by hour and considered as a point in the time series sequence. 
To evaluate the effect of other factors, such as weather conditions, on the future air pollution value, we use an NN model. The NN aims to capture the influence of weather factors and merge with the output of an LSTM model by an appropriate weight. The hybrid model is shown in Fig. 4.

\begin{figure}
    \centering
    \includegraphics[width=0.8\linewidth]{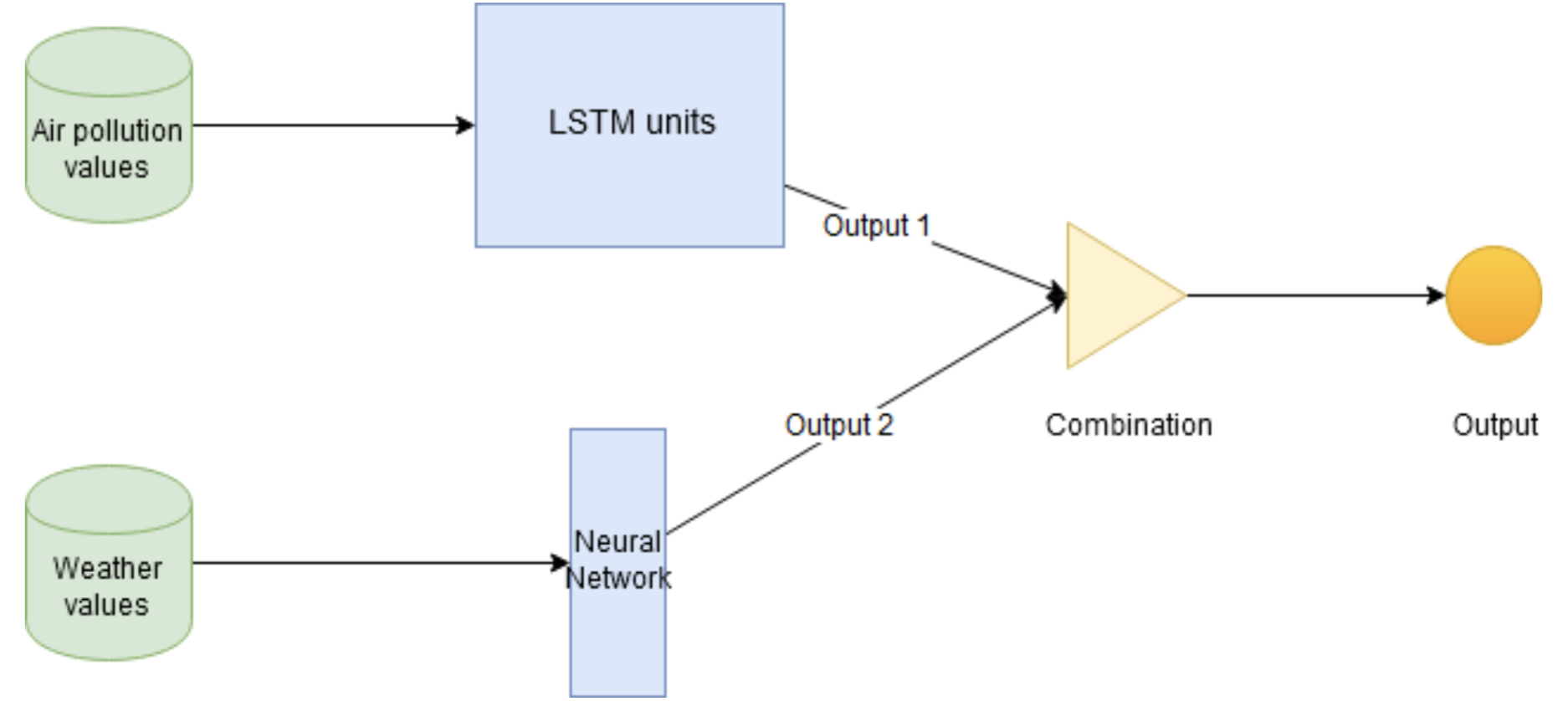}
    \caption{The hybrid model of LSTM units and a neural network model.}
    \label{fig:hybrid}
\end{figure}

The formula for this weighted combination model, as follows:
\begin{equation}
    y = \alpha \times y_\textrm{lstm} + (1 - \alpha) \times (w x_w + b),
\end{equation}

where $y$ is the final output, $y_\text{lstm}$ is the output of the LSTM unit, $x_w, w, b$ are the input, weight, and bias of the NN model, $\alpha$ is the weighted parameter, $\mathbf{0 \leq \alpha \leq 1}$.

\section{Experiments and Evaluation}

\subsection{CNN for predicting air pollution values in real-time}

\begin{figure}
    \centering
    \includegraphics[width=0.95\linewidth]{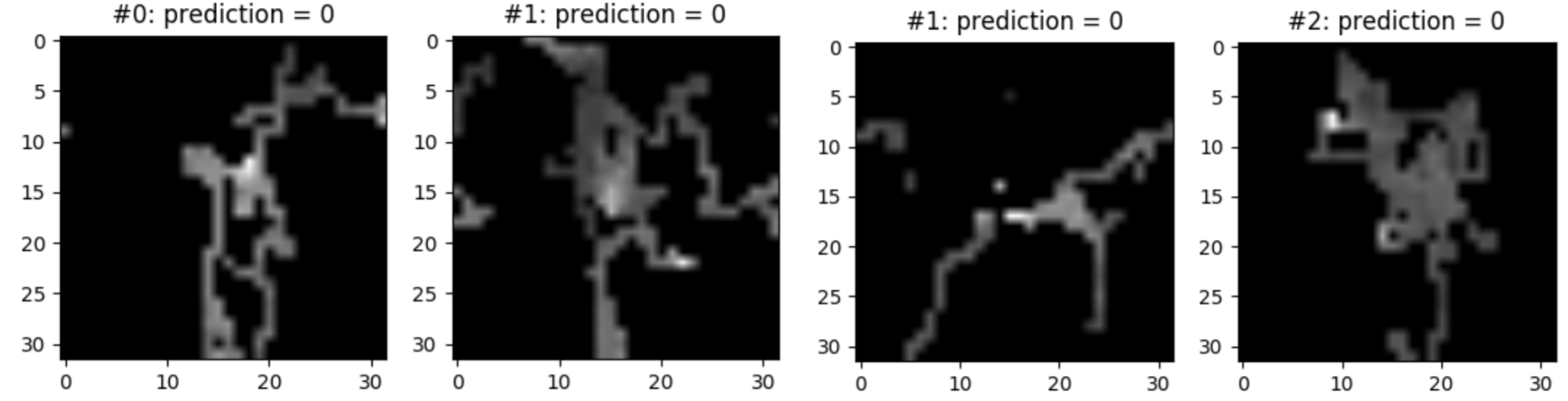}
    \caption{Prediction values with the testing set (01/2018). Prediction = 0 means outdoor air is Good (for health).}
    \label{fig:good_pred}
\end{figure}

The output of the previously described CNN model is to predict air pollution in 1 of 4 states: Good (0), Moderate (1), Unhealthy (2) and Hazardous (3). We use the training set is the data of sensor values in 4 months, from 09/2017 to 12/2017 and the testing set is the data of month 01/2018. All values are normalized to the range 0-1. The network is built by Tensorflow deep learning framework [11] and trained for 100 epochs with a batch size of 64 and using Adam for gradient optimizing. We got the testing accuracy approximate 74\%. For evaluation, because the real-time air pollution sensor dataset is unique, we do not have related works to compare our CNN model. Regarding real-time air pollution prediction, one can aggregate data in e.g. 1 minute each but with the interval fixed at 1 hour to get 1-hour data slice. Then we can apply the proposed CNN model to this new data and got prediction value in 1 of 4 states. Fig. 5 shows some of our results on the testing data.

\subsection{The hybrid model of LSTM and NN}
We also use the Tensorflow framework [11] to build a hybrid model of LSTM units and an NN model (weather conditions model). The training dataset is Daegu air pollution sensor data from 09/2017 to 01/2018 with 4 months 09~12/2017 as the training set and 1 month of 01/2018 as the testing set.
Recently, Recurrent Neural Network (RNN) and LSTM model are prevalent for time series prediction models, particularly air pollution values [12]. To evaluate our hybrid model with other related works, we make a standalone RNN model, a standalone LSTM model, and compare it with our hybrid model. The evaluation metric is defined as the relative mean absolute error (RMAE), as follows: 

\begin{equation}
    error = \frac{1}{m} \sum_{i=1}^m \frac{y_i - \hat{y_i}}{y_i},
\end{equation}

where $error$ is the evaluation error, $m$ is the number of evaluated examples, $y_i$ is a ground-true air quality value and $\hat{y_i}$ is a predicted air quality value.

Since our hybrid model is the combination of an LSTM model and an NN model with a weighted parameter value $\alpha$ ($0 \leq \alpha \leq 1$), we gauge the value $\alpha$ between 0.0 and 1.0 and compare the validation error of some baseline models and other weighted models corresponding to each $\alpha$. The results are shown in Table 1.

\begin{table}[ht]
\centering
\caption{Validation error of baseline models and our hybrid model with different $\alpha$}
\begin{tabular}{c >{\centering\arraybackslash}p{2.7cm} >{\centering\arraybackslash}p{2.7cm} >{\centering\arraybackslash}p{2.7cm} >{\centering\arraybackslash}p{2.7cm}}
\toprule
\textbf{$\alpha$} & \textbf{Standalone RNN model} & \textbf{Standalone LSTM model} & \textbf{RNN + NN model} & \textbf{Hybrid model (LSTM + NN)} \\
\midrule
1.0 & \textbf{5.010952} & 5.775150 & \textbf{5.010952} & 5.775150 \\
0.9 & & & 5.207657 & \textbf{4.656111} \\
0.8 & & & \textbf{4.574787} & 4.880526 \\
0.7 & & & 6.302503 & \textbf{4.895847} \\
0.6 & & & 4.946835 & \textbf{4.705889} \\
0.5 & & & 4.834772 & \textbf{4.257993} \\
0.4 & & & \textbf{4.265581} & 5.341873 \\
0.3 & & & \textbf{6.811509} & 7.887727 \\
0.2 & & & 5.673909 & \textbf{4.966643} \\
0.1 & & & 5.198360 & \textbf{4.546102} \\
0.0 & & & 6.597609 & \textbf{6.476847} \\

\hline
\end{tabular}
\end{table}

As in Table 1, when $\alpha$ is 1.0 we have a standalone RNN or LSTM model to predict time series data, and with $\alpha$ being 0, we have a standalone NN model for only weather condition factors. A smaller error indicates a better model (shown in bold).
We can infer from Table 1 that a standalone RNN model may be better than a standalone LSTM because its validation error is smaller. However, the number of smaller validation errors of Hybrid model (LSTM+NN) is more than that of RNN+NN model for 11 values of $\alpha$ from 0 to 1 (7/11 smaller validation errors compare to 4/11). This result indicates the combination of an LSTM model and an NN model is better than RNN + NN model. We think it is because of an LSTM unit could represent long time series better than an RNN.
Also, as in Table 1 experiment results, we could state that a combination of a time series model (RNN or LSTM) with an NN model is better than a standalone RNN or LSTM only because the validation errors of the combination model is smaller than a standalone model (see with $\alpha$ is 0.1, 0.2, 0.5~0.9). 

\section{Conclusion}
In this paper, we introduced a real-time air pollution sensor data collected in Daegu, Korea. The data volume is massive and in both spatial and temporal format. For spatial distribution air pollution, we proposed to consider the whole city as a grayscale image and applied a CNN model for real-time air pollution prediction. For temporal information, we present a simple but efficient hybrid model of an LSTM for time series data and an NN model for other impact factors, such as weather conditions. In the future, we will compare the prediction model based on this real-time sensor dataset with other hourly-collected ones from other cities in Korea and also around the world.

\section*{Acknowledgment}
This research was supported by 2018 grant of the Seoul Urban Data Science Lab Project (0660-20180009) through the Seoul Digital Foundation, funded by the Seoul Metropolitan Government.

\bibliographystyle{splncs04}

\end{document}